\documentclass[12pt]{article}

\input{tcilatex}

\begin{document}

\author{V.P.Akulov$^{\text{a,b}}$, S.Catto$^{\text{a,b,c}}$,\fbox{A.I.Pashnev} \and $%
^{\text{a}}$Department of Natural Science, Baruch College of the CUNY, New
York, \and NY 10010, USA\vspace{0.3cm} \and $^{\text{b}}$Physics Department,
\and The Graduate School and University Center the City University of N Y,
\and 365 Fifth Ave,New York, NY 10016-4309,USA \and $^{\text{c}}$Center for
Theoretical Physics, \and The Rockefeller University, \and 1230 York Ave,
New York, NY 10021-6399, USA \and $^{\text{d}}$Bogoliubov Laboratory of
Theoretical Physics, JINR , Dubna ,14980, \and Russia\vspace{0.5cm}}
\title{N=2 SuperTime Dependent Oscillator and Spontaneous Breaking of Supersymmetry}
\date{September 30 ,2004 }
\maketitle

\begin{abstract}
Using the nonlinear realizations of the N=2 superVirasoro group we construct
the action of the N=2 Superconformal Quantum Mechanics(SCQM) with additional
harmonic potential.We show that SU(1,1|1) invariance group of this action is
nontrivially embedded in the N=2 Super Virasoro group.The generalization for
the (super)time dependent oscillator is constructed.In a particular case
when the oscillator frequency depends on the proper-time anticommuting
coordinates the unusual effect of spontaneous breaking of the supersymmetry
takes place: the Masses of bosons and fermions can have different nonzero
values.
\end{abstract}

\section{Introduction}

The Time Dependent Oscillator ( so called Ermakov system\cite{E}) has so
many physical applications (see \cite{ACP1} for list of some references) and
was the subject of vigorous research for decades especially due to its
elegant mathematical properties and application potential of its invariant.

A vital modification of the Time Dependent Oscillator includes an additional
term in the potential proportional to the inverse square of the coordinate -
it is often referred to as the anharmonic oscillator. This extra term is
conformally invariant. Analogous Conformal Quantum Mechanics (CQM) was
investigated in detail by De Alfaro, Fubini and Furlan\cite{AFF}. It was
shown in their paper that the consistent quantum treatment of the model
assumes the transition to the new time coordinate which transpires to be
equivalent to the introduction of an additional oscillator-like term with
constant frequency in the potential. Therefore, the emerging physical
Hamiltonian represents the anharmonic oscillator with time independent
frequency $\omega$.

The most adequate approach for understanding the geometrical meaning of CQM
and SCQM is the method of nonlinear realizations of the symmetry groups
underlying both the theories - the group $SL(2,R)$ and its supersymmetrical
generalizations $SU(1,1|1)$, and $SU(1,1|2)$, respectively\cite{IKL1},\cite
{IKL2}. In spite of its power, this method does not allow the
oscil\-lator-like potentials introduced in \cite{AFF} to be included in the
Hamiltonian of the theory. As was shown in \cite{ACP1} the explanation for
this lies entirely in the fact that in the presence of the oscillator-like
term the invariance group of the nonsupersymmetric action, though being the
Conformal Group, is realized by more complicated transformations. These
transformations for the constant $\omega$, as well as for the time-dependent
one (Ermakov system), can naturally be embedded in the reparametrization
group of the time variable which is isomorphic to the centerless Virasoro
group. This embedding is rather nontrivial in the case of nonvanishing $%
\omega$.

As will be shown in this paper, the supersymmetrization of the model leads
to new posibility of supertime dependence of the oscillator frequency. As a
result, the masses of bosons and fermions can have different nonzero values.

The structure of the paper is as follows. In Section 2 we apply the
nonlinear realizations method to the Virasoro group and its three
dimensional subgroup $SL(2,R)$. Using the Cartan's invariant Omega-forms we
construct the action for Conformal Quantum Mechanics and describe the
mechanism for appearance of oscillator-like terms in the Omega-forms and,
correspondingly, in the action. We show how the symmetry group of this
action, $SL(2,R)$, is nontrivially embedded in the Virasoro group and
generalize these results to the Ermakov systems with timedependent
oscillator frequency.

In Section 3 we supersymmetrize the above constructions and illustrate the
mechanism of appearing different nonzero masses for bosons and fermions.

Some further anticipations of the formalism developed are included in the
conclusions.

\setcounter{equation}0

\section{Nonlinear Realization of the Virasoro Group}

The generators of the infinite dimensional reparametrization
(diffeomorphisms) group on the line parametrized by some parameter $s$ are $%
L_m=is^{m+1}\frac{d}{d s}$ and form the Virasoro algebra without central
charge 
\begin{equation}  \label{Virasoro}
\left[L_n,L_m\right]=-i(n-m)L_{n+m}.
\end{equation}
If one restricts oneself to the transformations regular at the origin $s=0$,
it is convenient to parametrize the Virasoro group element as\cite{IK1,IK2} 
\begin{eqnarray}  \label{coset1}
&&G=e^{i\tau L_{-1}} \cdot e^{ix_1L_1} \cdot e^{ix_2L_2} \cdot
e^{ix_3L_3}\ldots e^{i{x_0}L_0},
\end{eqnarray}
where all multipliers, except the last one, are arranged by the conformal
weight of the generators in the exponents.

The transformation laws of the group parameters $\tau, x_n$ in (\ref{coset1}%
) under the left action 
\begin{equation}  \label{left1}
G^{\prime}=(1+ia)G,
\end{equation}
where the infinitesimal element $a$ belongs to the Virasoro algebra 
\begin{equation}  \label{epsilon}
a = a_0 L_{-1}+{a_{1}}L_0+ {a_{2}}L_1+...+{a_{m}}L_m+... =
\sum_{n=0}^{\infty}a_nL_{n-1},
\end{equation}
are 
\begin{equation}  \label{Transtau}
\delta \tau =a(\tau),\quad \delta x_0=\dot{a}(\tau),\quad \delta {x_1}= -%
\dot{a}(\tau)x_1+ \frac{1}{2}\ddot{a}(\tau),\quad \delta {x_2}= -2\dot{a}%
(\tau)x_2+ \frac{1}{6}\stackrel{\ldots}{a}(\tau),
\end{equation}
where the infinitesimal function $a(\tau)$ is constructed out of the
parameters $a_n$ 
\begin{equation}  \label{a}
a(\tau)=a_0+a_1\tau+a_2\tau^2+ +a_3\tau^3\ldots =
\sum_{n=0}^{\infty}a_n\tau^n.
\end{equation}
One can see from (\ref{Transtau}) that the parameter $\tau$ transforms
precisely as the coordinate of the one-dimensional space under the
reparametrization. The parameters $x_0$ and $x_1$ transform,
correspondingly, as the dilaton and one-dimensional Cristoffel symbol. In
general, the transformation rule for $x_n$ contains the $(n+1)$-st
derivative of the infinitesimal parameter $a(\tau)$.

All parameters $x_n, n=0,1,2,...$ in (\ref{coset1}) in physical models are
considered as fields in one-dimensional space parametrized by the coordinate 
$\tau$.

The conformal group $SL(2,R) \sim SU(1,1)$ in one dimension is a
three-parameter subgroup of (\ref{coset1}), namely the one generated by $%
L_{-1}, L_0$ and $L_1$. Its group element is a product of the first two and
last one multipliers in expression (\ref{coset1}) 
\begin{equation}  \label{coset2}
G_C=e^{i\tau L_{-1}} \cdot e^{ix_1L_1} \cdot e^{i{x_0}L_0}.
\end{equation}
In other words, the $SL(2,R)$ group is embedded in the Virasoro group in the
simplest way by the conditions 
\begin{equation}  \label{Conditions}
x_n=0,\quad n\geq 2
\end{equation}
The infinitesimal transformation function $a(\tau)$ (\ref{epsilon}), which
conserves the conditions (\ref{Conditions}), contains only three parameters 
\begin{equation}  \label{3parameter}
a(\tau)= a_0+a_1\tau +a_2\tau^2.
\end{equation}

It is convenient to introduce new variables playing the roles of the
coordinate and momentum of the particle 
\begin{equation}  \label{newVar}
x=e^{x_0/2},\quad p=x_1 x,
\end{equation}
for which the conformal group infinitesimal transformations are 
\begin{equation}  \label{Transtau3parameter}
\delta \tau =a(\tau),\quad \delta x=\frac{1}{2}\dot{a}(\tau)x,\quad \delta {p%
}= -\frac{1}{2}\dot{a}(\tau)p+ \frac{1}{2}\ddot{a}(\tau)x,
\end{equation}
with $a(\tau)$ given in this case by (\ref{3parameter}).

Infinitesimal transformations (\ref{Transtau3parameter}) generate the
symmetry group of the Conformal Quantum Mechanics of \cite{AFF} with the
action 
\begin{equation}  \label{ActionAFF}
S = {\frac{1}{2}} \int d\tau \left( \dot{x}^2 - {\frac{\gamma}{x^2}} \right).
\end{equation}
As was shown in \cite{IKL1} (see also \cite{ACP}) this action can be
naturally described in terms of the invariant differential Cartan's form 
\begin{equation}  \label{CartanC}
\Omega_C=G_C^{-1} dG_C=\Omega_{-1}L_{-1}+\Omega_0 L_0+\Omega_1 L_1
\end{equation}
connected with the parametrization (\ref{coset2}) of the conformal group.
Explicit calculations give 
\begin{equation}  \label{Cbose}
\Omega_{-1}=\frac{d\tau}{x^2}, \quad \Omega_0= \frac{dx-pd\tau}{x},\quad
\Omega_1=xdp-pdx+p^2d\tau.
\end{equation}

All these differential forms are invariant under transformations (\ref
{Transtau3parameter}) and can be used for construction of an invariant
action. The simplest one is the combination linear in $\Omega$-forms\cite
{IKL1} 
\begin{equation}  \label{BoseAction}
S=-\frac{1}{2}\int \Omega_1-\frac{\gamma}{2}\int\Omega_{-1}= \frac{1}{2}\int
d\tau \left(-x {\dot p}+p{\dot x} - p^2- \frac{\gamma}{x^2} \right)~.
\end{equation}
The first term in this expression is appropriately normalized to get the
correct kinetic term. The parameter $\gamma$ plays the role of a
cosmological constant in one dimension because $\Omega_{-1}$, which
corresponds to the translation generator $L_{-1}$, is the differential $1$%
-form einbein.

Action (\ref{BoseAction}) is the first order representation of that
describing Conformal Mechanics of De Alfaro, Fubini and Furlan\cite{AFF}.
Indeed, one can find $p$ by solving its equation of motion, insert it back
in the Lagrangian and get the second order action (\ref{ActionAFF}).

From the point of view of underlying physics action (\ref{ActionAFF}) is not
a satisfactory one because the corresponding quantum mechanical Hamiltonian
does not have the ground state. The modification of this action with the
appropriate energy spectrum was considered in \cite{AFF}. It includes the
additional harmonic oscillator term 
\begin{equation}  \label{ActionAFFHarmonic}
S_1 = {\frac{1}{2}} \int d\tau \left( \dot{x}^2 - {\frac{\gamma}{x^2}}%
-\omega^2 x^2 \right).
\end{equation}
Though the action (\ref{ActionAFFHarmonic}) contains the dimensional
parameter $\omega$, it is invariant under transformations of the conformal
group realized by more complicated expressions, as we will see.

As we have already mentioned in Introduction, action (\ref{ActionAFFHarmonic}%
) cannot be described in the framework of nonlinear realizations of the $%
SL(2,R)$ group, parametrized as in (\ref{coset2}). Instead, we will consider
the embedding of this group in the Virasoro group (\ref{coset1}) by
conditions different from the simplest ones (\ref{Conditions}). The
structure of the component $\Omega_1^V$ in the Cartan's Omega-form connected
with the Virasoro group 
\begin{equation}  \label{CartanV}
\Omega_V=G^{-1} dG=\Omega_{-1}L_{-1}+\Omega_0 L_0+\Omega_1^V L_1+ \Omega_2^V
L_2+\ldots
\end{equation}
may serve as a hint in the choice of appropriate conditions. The components $%
\Omega_{-1}$ and $\Omega_{0}$ coincide with the corresponding components (%
\ref{Cbose}). Though the components $\Omega_2^V, \Omega_3^V, \ldots$ depend
in general on all parameters $x_n$, the component $\Omega_1^V$ 
\begin{equation}  \label{CartanV1}
\Omega_1^V=xdp-pdx+p^2d\tau-3x_2x^2d\tau
\end{equation}
depends, in addition to the phase space variables $(x,p)$, only on the
parameter $x_2=x_2(\tau)$. So the last term in expression (\ref{CartanV1})
is the only difference from the corresponding expression (\ref{Cbose})
calculated for representation (\ref{coset1}) of the $SL(2,R)$ group.
Moreover, if we take 
\begin{equation}  \label{x2omega}
x_2(\tau)=-\frac{1}{3}\omega^2,\quad \omega=const,
\end{equation}
we obtain exactly an oscillator-like term in the action 
\begin{equation}  \label{OmegaAction}
S=-\frac{1}{2}\int \Omega_1^V-\frac{\gamma}{2}\int\Omega_{-1}= \frac{1}{2}%
\int d\tau \left(-x {\dot p}+p{\dot x} - p^2-\omega^2x^2- \frac{\gamma}{x^2}
\right)~,
\end{equation}
which coincides with the action $S_1$ (\ref{ActionAFFHarmonic}) in the
second order form.

The component $\Omega_1^V$ (\ref{CartanV1}) is by construction invariant
under the arbitrary infinitesimal transformations (\ref{left1}) of the
Virasoro group The consistency condition of this transformation law with the
demand that $\omega=const$ can be written in the form 
\begin{equation}  \label{EqOmegaConst}
\stackrel{\ldots}{a}(\tau)+4\omega^2\dot{a}(\tau)=0.
\end{equation}
The solution of this differential equation gives 
\begin{equation}  \label{ParOmegaConst}
a(\tau)=a_0+a_1\sin(2\omega\tau)+a_2\cos(2\omega\tau).
\end{equation}
So the action of Conformal Mechanics (\ref{ActionAFFHarmonic}) with the
additional oscillator-like potential is invariant under the three parameter
transformation (\ref{ParOmegaConst}).

In general the variable $x_2(\tau)$ can be an arbitrary function of time.
Nevertheless, it cannot be a dynamical variable because, as one can easily
see from expression (\ref{CartanV1}), it plays the role of a Lagrange
multiplier leading to the equation of motion $x^2=0$\footnote{%
If the variable $x$ carries in addition some index $I$ - $x\rightarrow x_I$
the situation drastically changes when this index describes the vector
representation of the rotation group of the space-time with the signature $%
(D,2)$. In this case, the action is given by the sum of $D+2$ expressions (%
\ref{CartanV1}) (with the corresponding signs) and it describes the massless
particle in $D$ - dimensional space-time\cite{P2} (or the spinning particle
if instead of the Virasoro group one considers the reparametrization group
in the superspace $(1,N)$ with one bosonic and $N$ Grassmann coordinates\cite
{MAAP})} So instead of being the constant as in the previous Subsection, the
parameter $x_2$ in a physical model can be at most some \emph{fixed}
function $x_2(\tau)$. If we look for invariance transformations of action (%
\ref{OmegaAction}) with the time dependent frequency $\omega^2(\tau)$ ($%
x_2(\tau)=-\omega^2(\tau)/3$), it means that after the time transformation (%
\ref{Transtau}) $\tau\rightarrow \tau^{\prime}=\tau+a(\tau)$ the functional
dependence should remain the same: $x_2(\tau)\rightarrow x_2(\tau^{\prime}),
\delta x_2(\tau)=a(\tau)\dot x_2(\tau)$. The transformation law (\ref{Trans2}%
) leads then to the equation for the infinitesimal parameter $a(\tau)$ 
\begin{equation}  \label{EqOmegaTime}
\stackrel{\ldots}{a}(\tau)+4\omega^2(\tau)\dot{a}(\tau)+ 2\frac{d}{d\tau}{%
(\omega^2(\tau))}{a}(\tau)=0.
\end{equation}
This differential equation of the third order with the time dependent
coefficients has the solution in the form\cite{K} 
\begin{equation}  \label{Solution}
a(\tau)=C_1 u_1^2+C_2 u_1u_2+C_3 u_2^2,
\end{equation}
where $C_1, C_2, C_3$ are three infinitesimal constants, and the functions $%
u_1(\tau), u_2(\tau)$ form the fundamental system of solutions to the
auxiliary equation 
\begin{equation}  \label{Aux}
\ddot{u}(\tau)+\omega^2(\tau){u}(\tau)=0.
\end{equation}
For the time independent $\omega$ this solution reproduces the ones given by
(\ref{ParOmegaConst}). For different particular forms of $\omega^2(\tau)$
equation (\ref{EqOmegaTime}) becomes, for example, the Lame, Matieu, Hill
etc. equations \cite{K}, each playing very important role in physics.

So solution (\ref{Solution}) of equation (\ref{EqOmegaTime}) describes the
invariance transformations of the action for the Time Dependent Oscillator
with the frequency $\omega(\tau)$.

\setcounter{equation}0

\section{$N=2$ SuperTime dependent oscillator}

The $N=2$ SuperVirasoro algebra is formed by the Virasoro generators $L_n$,
complex supergenerators $G_r, {\bar G}_s$ and $U(1)$ Kac-Moody algebra
generators $T_k$. The indices $n,k$ and $r,s$ are arbitrary integer and
halfinteger numbers correspondingly. This algebra has the following form 
\begin{eqnarray}
\left[L_m,L_n\right]&=&-i(m-n)L_{m+n},\quad \left[L_m,T_k\right]=ikT_{m+k} 
\nonumber \\
\left[L_m,G_s\right]&=&-i(\frac{m}{2}-s)G_{m+s},\quad \left[L_m,{\bar G}%
_s\right]=-i(\frac{m}{2}-s){\bar G}_{m+s} \\
\left[T_k,G_s\right]&=&-\frac{i}{2}G_{k+s},\quad \left[T_k,{\bar G}_s\right]=%
\frac{i}{2}{\bar G}_{k+s}  \nonumber \\
\{G_r,{\bar G}_s\}&=&-2L_{r+s}-2(r-s)T_{r+s}.  \nonumber
\end{eqnarray}

If one restricts to the subalgebra with indices in the region $m,n \geq
-1,k\geq 0, r,s\geq -1/2$, which we will call in what follows as $N=2$
Superconformal Algebra (SCA), it is convenient to parameterize the coset
space of the corresponding group over the $U(1)$ subgroup generated by $T_0$
in the following form \cite{IK1,IK2} 
\begin{eqnarray}
G_{\mathcal{I}}\!\!\!\!&=&\!\!\!\!e^{i\tau L_{-1}} \cdot e^{\bar\theta
G_{-1/2}+ \theta {\bar G}_{-1/2}} \cdot e^{\bar\psi G_{1/2}+\psi{\bar G}%
_{1/2}} \cdot e^{iU^{(1)}L_1}\cdot e^{V^{(1)}T_1} \cdot  \nonumber \\
\!\!\!\!&&\!\!\!\! e^{\bar\Theta^{(3/2)}G_{3/2}+ \Theta^{(3/2)}{\bar G}%
_{3/2}} \cdot e^{iU^{(2)}L_2} e^{V^{(2)}T_12} \cdot \,\cdots \,\cdot
e^{iU^{(0)} L_0}.  \label{cosetn2}
\end{eqnarray}
where all multipliers, except the last one, are arranged by the conformal
weight of the generators in the exponents. The transformation laws of the
group parameters $\tau$, $\theta$, $\psi$, $U^{(0)}$, $U^{(1)}$, $V^{(1)}$, $%
...$ in (\ref{cosetn2}) under the left action 
\begin{equation}  \label{Trans2}
G^{\prime}=(1+ia)G,
\end{equation}
where infinitesimal element $a$ belongs to the $N=2$ Virasoro algebra, are
incoded in the infinitesimal real superfunction 
\begin{equation}  \label{Transn2}
\Lambda=a(\tau)+2i\theta\bar\epsilon(\tau)+2i\bar\theta\epsilon(\tau)+\theta%
\theta b(\tau)
\end{equation}
and are, in particular, 
\begin{eqnarray}  \label{Transtaun2}
\delta \tau &=&\Lambda-\frac{1}{2}(\theta D+\bar\theta{\bar D})\Lambda, \\
\delta \theta&=&-\frac{i}{2}{\bar D}\Lambda,  \label{Transtheta} \\
\delta \bar\theta&=&-\frac{i}{2}{D}\Lambda,  \label{Transthetabar} \\
\delta V^{(1)}&=& -\dot{\Lambda}V^{(1)}- \frac{1}{2}(D{\bar D}-{\bar D}D)%
\dot{\Lambda},  \label{TransV1} \\
\end{eqnarray}
where $D$ and ${\bar D}$ are supercovariant derivatives $D=\partial/\partial%
\theta+i\bar\theta\partial/\partial\tau,$ ${\bar D}=\partial/\partial\bar%
\theta+i\theta\partial/\partial\tau$.

Parameters $\tau, \theta$ and $\bar\theta$ transform as coordinates of $N=2$
superspace. All other parameters can be wieved as superfunctions in this
superspace.

As was shown in \cite{P2}, the superspace action constructed with the help
of Cartan's invariant forms (we omit here details, as well, as
supersymmetrization of the conformal potential $1/x^2$) has the form 
\begin{equation}  \label{N2action}
S=\int \mathrm{d}\tau\;\underline{\mathrm{d}\theta}\; \underline{\mathrm{d}%
\bar\theta} (\frac{{1}}{2} DX \bar{D} X + \frac{1}{4}V^1 X^2),
\end{equation}
where $X=e^{U^{(0)}}= x(\tau)+\mathrm{i}\bar\theta\gamma(\tau)+ \mathrm{i}%
\theta\bar\gamma(\tau)+\bar\theta\theta F(\tau)$ is the $N=2$ superfield
coordinate.

The last term in (\ref{N2action}) after superspace integration reproduces
the mass terms for bosonic $x(\tau)$ and fermionic $\gamma(\tau),\bar\gamma(%
\tau)$ coordinates. If the function $V^1$ is a constant, both masses are
equal. On the other hand, $V^1$ can depend on all coordinates of the
superspace. The consistency equation analogous to (\ref{EqOmegaTime}) have
now the form 
\begin{equation}
\Lambda{\dot V^1}-i/2 D\Lambda {\bar D}V^1-i/2 {\bar D}\Lambda
DV^1+\dot\Lambda V^1+ 1/2(D{\bar D}-{\bar D}D)\dot\Lambda=0.
\end{equation}
The solutions of this equation define the embedding of the $SU(1,1|1)$
invariance group of the action in the $N=2$ Super Virasoro group.

As an example, one can take $V^1=2\omega_1+2\bar\theta\theta\Delta$. Then
masses of boson and fermions will be different 
\[
M_f=\omega_1,\quad M_b^2=\omega_1^2+\Delta. 
\]

\section{Conclusions}

In this paper we applied the methods of nonlinear realizations approach for
construction of the actions of SuperConformal Quantum Mechanics, as well, as
the action of the SuperTime Dependent Oscillator. Actions are invariant
under the transformations of $SU(1,1|1)$, which is nontrivially embedded in
the $N=2$ Super Virasoro group. It would be interesting to carry up the
analogous considerations in the more complicated theories, such $N=4$
SuperConformal Quantum Mechanics. The mechanism of spontaneous breakdown of
the supersymmetry established in this paper gives the possibility to
regulate boson and fermion masses without destroing the symmetry properties
of the model. The question whether this mechanism is specific only to one
dimensional models or the same mechanism will also be work for D= 4
dimensional models -is still an open.\quad

\section*{Acknowledgements}

A.P. thanks the members of Graduate School and University Center, the City
University of New York, where the essential part of this work was done. The
work of A.P. was partially supported by INTAS grant, project No 00-00254,
RFBR grant, project No 03-02-17440 and a grant of the Heisenberg-Landau
program.

\medskip This is a last paper of our dear friend Anatoly Pashnev, a trully
beautyful human being and high level scientist who suddenly passed away on
March 30, 2004.

\end{document}